\documentclass[aps,prx,reprint,amsmath,amssymb,superscriptaddress, nofootinbib, 10pt]{revtex4-2}
\usepackage[english]{babel}
\usepackage{amsmath,amssymb,graphicx,color,comment}
\usepackage[pdfusetitle,bookmarks=true,colorlinks,citecolor=blue,urlcolor=blue, linkcolor=blue]{hyperref}
\usepackage[normalem]{ulem}

\usepackage{braket}
\usepackage{subfigure}
\usepackage{extarrows}
\usepackage{float}

\usepackage{mathtools}
\usepackage{adjustbox}

\usepackage{dcolumn}   
\usepackage{bm}        
\usepackage{amsfonts}  
\usepackage{lineno}

\newcommand{\exv}[1]{\left\langle #1 \right\rangle}
\newcommand{\pwisein}{\left\{ \begin{array}{ll}}
	\newcommand{\pwiseout}{\end{array}\right.}

\newcommand{\Ham}[0]{\mathcal{H}}

\newcommand{\eq}[1]{Eq.~(\ref{#1})}

\newcommand{\fc}[1]{({#1})}
\newcommand{\figc}[2]{Fig.\thinspace{}\ref{#1}\thinspace{}\fc{#2}}
\newcommand{\figcc}[3]{Fig.\thinspace{}\ref{#1}\thinspace{}\fc{#2} and \fc{#3}}
\newcommand{\figcs}[3]{Fig.\thinspace{}\ref{#1}\thinspace{}\fc{#2} - \fc{#3}}

\newcommand{\App}[1]{Appendix \ref{#1}}
\newcommand{\Secc}[1]{Section \ref{#1}}
\newcommand{\Reff}[1]{Ref.~\cite{#1}}

\usepackage{times}

\newcommand{\titleinfo}{Scattering Theory of Chiral Edge Modes in Topological Magnon Insulators}

\begin{document}

\title{\titleinfo}

\author{Stefan Birnkammer}%
\affiliation{Department of Physics, Technical University of Munich, 85748 Garching, Germany}
\affiliation{Munich Center for Quantum Science and Technology (MCQST), Schellingstra{\ss}e 4, 80799 M{\"u}nchen, Germany}
\author{Michael Knap}
\affiliation{Department of Physics, Technical University of Munich, 85748 Garching, Germany}
\affiliation{Munich Center for Quantum Science and Technology (MCQST), Schellingstra{\ss}e 4, 80799 M{\"u}nchen, Germany}
\author{Johannes Knolle}
\affiliation{Department of Physics, Technical University of Munich, 85748 Garching, Germany}
\affiliation{Munich Center for Quantum Science and Technology (MCQST), Schellingstra{\ss}e 4, 80799 M{\"u}nchen, Germany}
\affiliation{Blackett Laboratory, Imperial College London, London SW7 2AZ, United Kingdom}
\author{Alexander Mook}
\affiliation{Department of Physics, Johannes Gutenberg Universität Mainz, 55128 Mainz, Germany}
\author{Alvise Bastianello}
\affiliation{Department of Physics, Technical University of Munich, 85748 Garching, Germany}
\affiliation{Munich Center for Quantum Science and Technology (MCQST), Schellingstra{\ss}e 4, 80799 M{\"u}nchen, Germany}

\begin{abstract}
Topological magnon insulators exhibit robust edge modes with chiral properties similar to quantum Hall edge states.
However, due to their strong localization at the edges, 
interactions between these chiral edge magnons can be significant, as we show in a model of coupled magnon-conserving spin chains in an electric field gradient.
The chiral edge modes remain edge-localized and do not scatter into the bulk, and we characterize their scattering phase: for strongly-localized edge modes we observe significant deviation from the bare scattering phase. 
This renormalization of edge scattering can be attributed to bound bulk modes resonating with the chiral edge magnons, in the spirit of Feshbach resonances in atomic physics. 
We argue that the scattering dynamics can be probed experimentally with a real-time measurement protocol using inelastic scanning tunneling spectroscopy.
Our results show that interaction among magnons can be encoded in an effective edge model of reduced dimensionality, where the interactions with the bulk renormalize the effective couplings. 
Our work introduces a systematic way to determine the many-body effective theory for edge states in topological magnon insulators. 
\end{abstract}

\date{\today}
\maketitle
\section{Introduction}
Quantum magnets with a topologically \textit{trivial} ground state supporting topologically \textit{nontrivial} spin excitations are an emerging platform to study the topology of collective excitations~\cite{McClarty2022}. They are found in ordered magnets~\cite{Katsura2010, Onose2010, Zhang2013, Hoogdalem2013, Mook2014, Mook2014edge, Owerre2016a, Kim2016Kane, Li2016Weyl, Mook2016Weyl, Mook2019Coplanar, Mook2020hinge,Karaki2023} and quantum paramagnets~\cite{Romhanyi2015, McClarty2017topotriplon, Joshi2019, Bhowmick2021Weyltriplons,d2024kitaev}.
When time-reversal symmetry is broken, chiral spin excitations propagating along the boundaries of the sample give rise to the idea of topological magnonics~\cite{Shindou13, Shindou13b, Shindou14, Wang2017, Wang2018}. 
Although direct experimental evidence for chiral edge spin excitations is yet to be achieved, transport~\cite{Onose2010, Ideue2012, Hirschberger2015,czajka2023planar} and neutron scattering data~\cite{Chisnell2015, Chen2018CrI3, Chen2021, zhu2021topological,dos2023topological} compatible with topological spin excitations have been reported. However, the stability against decay of topological magnon modes still needs to be firmly established~\cite{McClarty2018, Koyama2023FlavorWaveEdgeBreakdown, habel2024}. The properties of  magnon edge modes can be investigated by inelastic scanning tunneling spectroscopy (STS) through local excitation~\cite{Feldmeier2020, konig_2020}. 
While much of the focus thus far has been on investigating the properties of isolated edge excitations, many experimental setups operate in a regime where multiple edge modes are excited simultaneously. For example, spintronics or general transport experiments investigate spin and energy transport governed by an ensemble of edge modes with various energies and velocities. Naturally, they will interact with one another and consequences of interactions are expected to be reflected in transport properties of the system. It is therefore an important open challenge to study the characteristics of interactions between edge modes.\\
\begin{figure}
    \centering
    \includegraphics[width=\columnwidth]{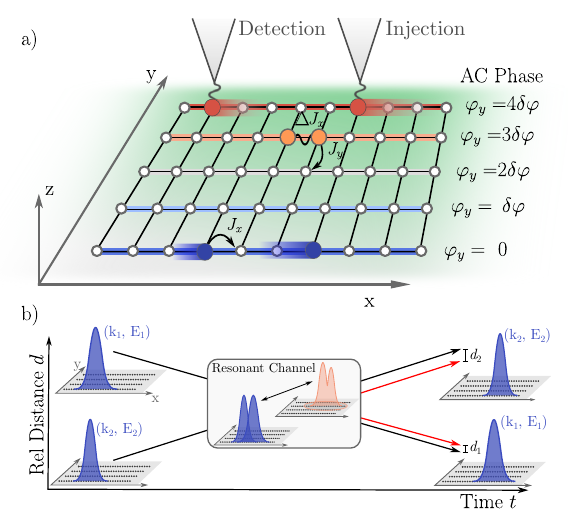}
    \caption{\textbf{Topological magnons in coupled wires.}
    a) Sketch of the model \eqref{eq:Hamiltonian-Chain}: we consider a two-dimensional system made of an array of spin$-1/2$ chains. The presence of a linear Aharonov-Casher (AC) phase induced by a constant electric field gradient in the $y$-direction $\vec{E} = \Delta E \vec{y}$, gives rise to localized bulk modes and chiral edge modes.
    b) Cartoon scattering of two chiral magnons at the same edge. Our results demonstrate that the scattering of edge modes is elastic (no scattering into the bulk), but resonances with bulk modes strongly renormalize effective interactions, manifesting in scattering shifts $d_{1}, d_{2}$ of the individual edge modes.}
    \label{fig:1}
\end{figure}
In this work, we consider a two-dimensional lattice of spins in which edge magnons are induced by the Aharonov-Casher (AC) effect \cite{Liu2011, Meier2003, Nakata2017b, Nakata2017AFM, Aharonov1984, Mook2018duality}; see \figc{fig:1}{a}. We use this tunable toy model as a proof of principle to develop a protocol for constructing an effective interacting edge Hamiltonian, thereby highlighting the peculiar role of resonances with bulk excitations for renormalization of interactions at the edge. 
We show that scattering of edge magnons remains confined to the edge with no leakage into bulk degrees of freedom, and we capture their dynamics within a two-body effective edge Hamiltonian
\begin{equation}\label{eq:eff_H_main}
\mathcal{H}_{\text{2-eff}}=\varepsilon(k_1)+\varepsilon(k_2)+U_{\text{2-eff}}(k_1,k_2)\, ,
\end{equation}
where $k_{1,2}$ are the edge-momenta of the magnons, and  $\varepsilon(k)$ is the single-magnon energy. The effective interaction $U_{\text{2-eff}}(k_1,k_2)$ is determined by the scattering phase $\phi(k_1,k_2)$ \cite{Korepin1993,Khokhlov2006,Kukulin1990}, which encodes the additional phase a two-magnon wavefunction acquires from scattering. Interestingly, we find $\phi(k_{1},k_{2})$ to be noticeably renormalized by virtual processes with bi-magnon bulk bound states.

Our work paves the way for a systematic development of an interacting theory of chiral edge magnons, and it provides the groundwork to use powerful theoretical methods of one-dimensional quantum systems to deepen our understanding of two-dimensional topological magnets.
Our results are pertinent for experimental studies: we discuss how scattering of chiral magnons can be explored using inelastic STS through a real-time injection measurement protocol, as shown in \figc{fig:1}{a}, where the presence of a non-trivial scattering phase manifests as a detectable shift of excitations after scattering; see \figc{fig:1}{b}.
Furthermore, understanding edge interactions could prove instrumental for future coherent magnon-based quantum technologies, such as magnonic interferometers and quantum circulators.\\
\section{The model}
\begin{figure}
    \centering
    \includegraphics[width=\columnwidth]{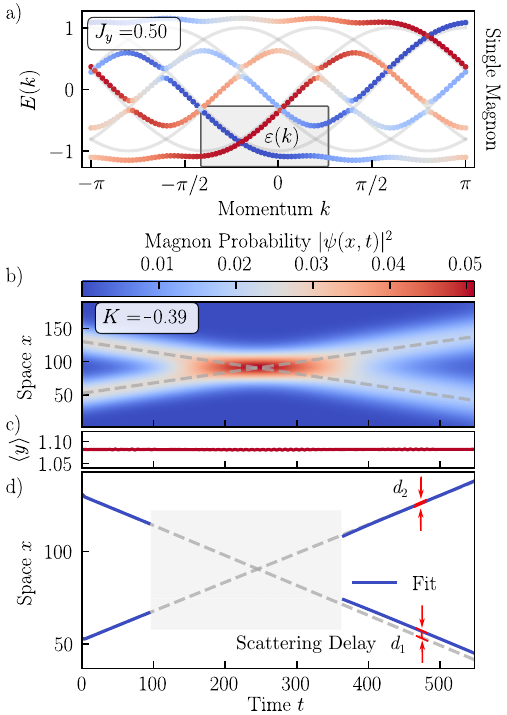}
    \caption{\textbf{Real time scattering of chiral magnons.} a) Single magnon spectrum with $\delta\varphi=2\pi/5$ for the case of uncoupled wires, consisting of multiple copies of single-wire bands shifted in the momentum space (gray lines). Inter-wire hopping $J_y\ne 0$ causes band repulsion, and the spectrum shifts to stationary bulk modes and chiral edge bands (within gray box). Location of the individual bands is indicated by color coding. 
    b) Real-time simulation of a scattering event of two chiral magnons with total momentum $K=-\frac{\pi}{8}$ and relative momentum $q=-\frac{\pi}{16}$, and parameters $(J_{y}, \Delta)=(0.5, 0.4)$. We track the bulk-integrated magnon distribution $\vert\psi(x)\vert^2$, with $x$ the relative distance.
    For a better visualization, we consider a comoving frame with the average velocity of the pair of magnons.
     The gray dashed lines are the trajectories of the initial position of the wavepackets, evolved with the initial velocities (free evolution).
    c) The state remains localized at the edge during the scattering, as shown in the average position in the transverse direction $\exv{y}$, where $\exv{y}\!=\!1$ corresponds to magnons being located in the lowest chain.
    d) We compare the free evolution (gray dashed line) with the peak positions (blue line), showing displacements $d_{1}, d_{2}$ after scattering.}
    \label{fig:2}
\end{figure}

We consider a ferromagnet with spin-conservation, ensuring magnons are stable excitations, and assume that any deviation from this condition can be neglected on the time scale of scattering experiments. As a toy model, we build a two-dimensional system from exactly solvable one-dimensional XXZ spin chains with tunable magnon-magnon interactions~\cite{Heisenberg1928, Bethe1931,takahashi2005thermodynamics}. The individual chains are arranged in the $x$-$y$ plane, aligned parallel in the $x$-direction, and labeled according to their equidistant positions along the $y$-axis, see \figc{fig:1}{a}.
The Hamiltonian for the $y$-th chain is
\begin{align}
\label{eq:Hamiltonian-Chain}
    \Ham^{(y)} = - \frac{J_{x}}{2}\sum_{x}\big( e^{-i \varphi_{y}}\sigma_{x,y}^{+} \sigma_{x+1,y}^{-}  + \mathrm{h.c.} \big) + \Delta \sigma_{x,y}^{z} \sigma_{x+1, y}^{z}. 
\end{align}
The magnons' interactions are controlled by the anisotropy $\Delta$, with magnons behaving as hardcore bosons in the limit $\Delta=0$. Next we stabilize a magnetically ordered state along the $z$-direction for all $\Delta$ via an external magnetic field. Chiral edge magnons can then be induced on top of this ordered state via an Aharonov-Casher (AC) phase~\cite{Liu2011, Meier2003, Nakata2017b, Nakata2017AFM, Aharonov1984, Mook2018duality}: an electric field gradient along the $y$-direction causes a chain-dependent phase $\varphi_{y} \propto \int_{(x, y)}^{(x+1, y)} \mathrm{d}\vec{x} \cdot (\vec{E} \wedge \hat{e}_{z}) \equiv y \delta\varphi$ for the intra-chain hopping. We measure energy scales in units of the intra-wire hopping and set $J_x\!=\!1$. The eigenstates of \eq{eq:Hamiltonian-Chain} have a definite number of magnons. 
In the absence of interchain coupling, the phase $\varphi_y$ is equivalent to different gauges of lattice momenta, resulting in a global shift of the one-magnon bands $k\to k+\varphi_y$, see \figc{fig:2}{a} (gray lines).
In the full Hamiltonian,
\begin{align}
\label{eq:Hamiltonian}
    \Ham = \sum_{y} \Ham^{(y)}  - J_{y} \sum_{x, y}\frac{1}{2} \Big(\sigma_{x,y}^{+} \sigma_{x,y+1}^{-}  + \mathrm{h.c.} \Big), 
\end{align}
the inter-wire hopping $J_y$ causes level repulsion between the wires' energy bands, opening up a bulk gap in the system. Only the energy branches associated with the edge wires escape level repulsion in certain momenta regions $k\in[k_1, k_2]$, giving rise to localized chiral edge modes with dispersion $\varepsilon(k)$, and group velocity $v(k)\!\equiv\!\partial_k \varepsilon(k)$ \cite{Meng2020}.
In the thermodynamic limit, bulk modes become localized and have flat energy bands, while edge modes have a well-defined chirality given by the sign of the group velocity, and they are dispersive, i.e. $\partial_k v(k)\!\ne\!0$, thus they scatter. We emphasize that also Ising inter-wire interactions can be added to \eq{eq:Hamiltonian}. While this leads to no qualitative changes in the physics it complicates the following discussion of bulk resonances and is hence only examined in \App{app:IsoInteractions}.\\
We show an example of the chiral spectrum in \figc{fig:2}{a} for inter-wire hopping strength $J_{y}\!=\!0.5$, and  we indicate the location of individual magnon modes in the bulk by color coding according to the sketch of \figc{fig:1}{a}, for $\delta\varphi=\frac{2\pi}{5}$. The chiral dispersion $\varepsilon(k)$ is located within the gray box. We fix the AC phase to this value throughout the rest of this work, but other values are straightforward and lead to similar results.\\
\section{Real-Time Scattering of Chiral Magnons} 
When scattering channels into bulk modes or other edge bands are closed, the scattering remains effectively one-dimensional. Due to energy and momentum conservation, two-magnon scattering is elastic preserving the individual momenta, but it remains highly non-trivial as interactions induce \emph{scattering shifts} $d_{1}, d_{2}$, displacing the magnons from their free trajectories.
This delay quantifies the strength of interactions between excitations.
We investigate this effect in real-time scattering of edge modes using exact diagonalization in the two magnon subspace. 
For this, we implement an initial state of two localized magnons of given total momentum $K$ and relative momentum $q$ separated from each other in the lattice. To get optimal overlap of our initial state with the eigenstates of the system, we construct the real space wavefunction from the single magnon eigenstates $\bar{\psi}_{k_{1}}(y_{1}), \bar{\psi}_{k_{2}}(y_{2})$ of the Hamiltonian with momenta $k_{1} = K/2 + q$ and $k_{2}=K/2 - q$. In real space, single magnon eigenstates are given by plane waves of momentum $k_{1}$ and amplitude $\bar{\psi}_{k_{1}}(y_{1})$ in wire $y_{1}$. We localize both single magnon states by sufficient dressing with Gaussian filters. The overall wavefunction $\psi_{k_{1},k_{2}}(x_{1}, x_{2}, y_{1}, y_{2})$  is hence given as a superposition of two single magnon states, i.e.
\begin{align}
    \psi_{k_{1}, k_{2}}(x_{1}, x_{2}, y_{1}, y_{2}) \propto\;  
    &\bar{\psi}_{k_{1}}(y_{1})\bar{\psi}_{k_{2}}(y_{2}) G_{\sigma X_{1}}(x_{1}) G_{\sigma X_{2}}(x_{2})\nonumber\\ 
    &\times e^{-i(k_{1}x_{1} + k_{2}x_{2})} \nonumber \\
    \text {with} \quad G_{\sigma X_{0}}(x)& \equiv \dfrac{1}{\sqrt{2\pi\sigma^2}}e^{-\frac{1}{2\sigma^{2}}(x - {X}_{0})^{2}}.  
\end{align}
Scattering between the individual magnon wavepackets initially localized at positions $X_{1}, X_{2}$ with width $\sigma$ is then studied using standard time evolution under the Hamiltonian  of \eq{eq:Hamiltonian}.
The results of an exemplary evolution are summarized in \figcs{fig:2}{b}{d}. The bulk-integrated magnon distribution $\vert\psi(x)\vert^2$ as a function of time is shown in \figc{fig:2}{b}. To confirm the absence of leakage into bulk modes or transitions to other edge states during scattering, we compute the wave packets' average edge localization $\exv{y}$, which stays constant in time, see \figc{fig:2}{c}.
In \figc{fig:2}{d}, we quantify the scattering shift $d_{1}, d_{2}$ by comparing the actual peak positions (solid blue line) with those predicted for free propagation (dashed gray line). For details on the numerical implementation see \App{app:CharacterizationManyBodyEdge}. 
While the scattering shift observed in real-time evolution is a direct manifestation of interactions, a more systematic study is provided by the scattering phase $\phi(k_1,k_2)$ which is directly connected with the scattering shift as $d_{1,2}=\partial_{k_{1,2}}\phi(k_1,k_2)$ \cite{Wigner1955}.\\
\begin{figure*}[t!]
    \centering\includegraphics[width=\textwidth]{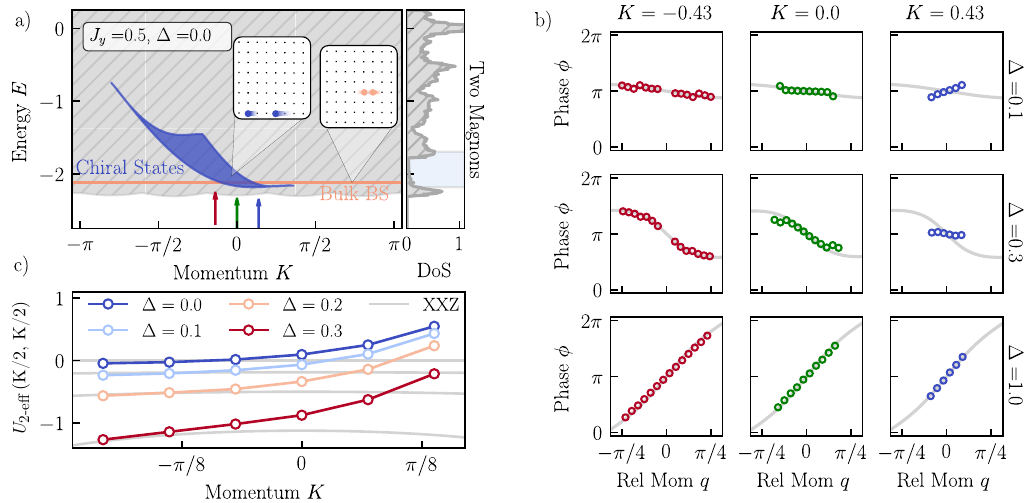}
    \caption{\textbf{Scattering phases.} a) Left: Allowed energy-momentum window for scattering magnons  at the lower edge (blue) embedded in the two-magnon continuum (gray area). The asymptotic energy region (blue) overlaps with non-dispersive bulk bands associated to bound states of two neighboring magnons (orange). The parameters are $(J_{y}, \Delta)\!=\!(0.5, 0.0)$.
    Right: Momentum-integrated two-magnon density of states capturing the low density of states within the bulk gap (lightblue shaded region). 
    b) Extracted scattering phases for asymptotic states of different relative momenta $q$ for fixed total momenta $K\in\{-0.43, 0, 0.43\}$ (left, middle, right column) for selected $\Delta\in\{0.1, 0.3, 1.0\}$. The one-dimensional XXZ prediction (gray line) is provided for comparison, highlighting a strong renormalization for positive $K$ and small $\Delta$.
    c) Comparison of equal-momenta effective interaction strength $U_\text{2-eff}(K/2,K/2)$ extracted from the scattering phase and chiral dispersion $\varepsilon(k)$ for numerically extracted results and analytical predictions for one-dimensional XXZ (gray lines).}
    \label{fig:3}
\end{figure*}

\section{Characterization of Edge Scattering Phases}
\label{ssec:EffectiveTheory}
When two wavepackets of edge magnons collide, they accumulate a scattering phase $\phi(k_1,k_2)$ depending on the asymptotic momenta $k_{1,2}$. For non-interacting particles, the scattering phase $\phi(k_1,k_2)$ vanishes; for hard-core bosons, $\phi(k_1,k_2)$ approaches $\pi$; and for genuinely interacting cases, it exhibits a non-trivial momentum dependence.
The scattering phase determines the effective interactions $U_{\text{2-eff}}(k_1,k_2)$ in Eq.~\eqref{eq:eff_H_main} through the inverse scattering method \cite{Korepin1993,Khokhlov2006,Kukulin1990}
: the solution is unique, and can be numerically found evaluating the Gel’fand-Levitan-Marchenko's equations~\cite{Khokhlov2006} or solving an optimization problem based on chosen trial potentials~\cite{Kukulin1990}. 
Instead of aiming for a full characterization of the effective interactions we can, however, focus on the emergent theory in the assumption that the momenta of the scattering magnons belong to a small shell $k\in I_p\equiv [p-\delta p, p+\delta p]$. In this approximation, we Taylor expand the dispersion law 
\begin{equation}
\varepsilon(k_{1,2})\simeq \varepsilon(p)+v(p)(k_{1,2}-p)+\frac{1}{2 m^{*}(p)}(k_{1,2}-p)^2 + \cdots, 
\end{equation}
where we introduced the velocity $v(k)= \partial_k \varepsilon(k)$ and effective mass $m^*(k)\equiv [\partial_k^2 \varepsilon(k)]^{-1}$, and terms of order $\mathcal{O}\big((k_{1,2}-p)^3\big)$ are not shown. Next, we approximate the interaction $U_{2-\text{eff}}(k_1,k_2)\simeq U_{2-\text{eff}}(p,p)$. Within this approximation, it is easy to convert $H_{\text{2-eff}}$ into a real space Hamiltonian, where the spatial coordinates $x_{1,2}$ conjugated to the small momenta shift $k_{1,2}-p$ are approximated as continuous
\begin{align}
\label{eq:eff_H_cont}
\Delta \mathcal{H}_{\text{2-eff}}\simeq&
-\tfrac{1}{2m^*(p)}(\partial^2_{x_1}+\partial^2_{x_2})+i v(p)(\partial_{x_1}+\partial_{x_2}) \nonumber \\ &+ c(p)\delta(x_1-x_2) \, ,
\end{align}
where $\Delta \mathcal{H}_{\text{2-eff}}= \mathcal{H}_{\text{2-eff}}-2\varepsilon(p)$ and $c(p)\equiv U_{\text{2-eff}}(p,p)$.
The Hamiltonian \eqref{eq:eff_H_cont} describes the two-particle sector of the Lieb-Liniger (LL) model~\cite{Lieb1963, Lieb1963a}: with a contact potential, in a moving frame with velocity $v(p)$. The scattering phase of this model is~\cite{Lieb1963, Lieb1963a} 
\begin{equation}
\phi_{\mathrm{LL}}(k_1,k_2)=-2\arctan\left(\tfrac{k_1-k_2}{m^{*}(p)c(p)}\right).
\end{equation}
By comparing the true scattering phase with the effective one, we can determine the effective interaction. Since we are within a small relative momentum approximation, we fix the effective interaction $c(p)$ through the scattering length $a(p)\equiv \lim_{k_1\to p}\partial_{k_1} \phi(k_1,p)$, and the LL correspondence $a(p)=-1/(m^{*}(p) c(p))$.
This, in particular, implies $U_{\text{2-eff}}(k,k)= -1/[m^*(k) a(k)]$.\\ 
Quantifying the effective theory at the edge hence requires characterization of the scattering phase $\phi(k_1, k_2)$. As an alternative to the real-space scattering protocol presented above, 
we systematically probe the scattering phase directly through the eigenstates of the two-dimensional problem, by searching the spectrum for scattering states of chiral edge modes. In the asymptotic region $x_1\ll x_2$, with $x_{1,2}$ the position of the magnons in the edge direction, the eigenfunction reads
\begin{align}
\psi(x_1,x_2,y_1, y_2)&\propto e^{ix_1 k_1+ix_2 k_2} \bar{\psi}_{k_1}(y_1)\bar{\psi}_{k_2}(y_2) \nonumber \\
& + e^{i\phi(k_1,k_2)}e^{i x_1k_2+ix_2 k_1}\bar{\psi}_{k_1}(y_2)\bar{\psi}_{k_2}(y_1),
\end{align}
where $e^{ikx}\bar{\psi}_{k}(y)$
is the one-magnon wavefunction.\\
In practice, we focus on the two-magnon sector of the Hilbert space, consider the center-of-mass (COM) frame and numerically diagonalize the Hamiltonian for each total momentum $K$ with standard numerical methods, further detailed in the \App{app:ExtractPhi}.
An example of the momentum-resolved low-energy continuum spectrum is shown in \figc{fig:3}{a} (gray shaded region) for parameters $(J_{y}, \Delta)\!=\!(0.5,0)$. The topological bulk gap is visible in the momentum-integrated density of states (right panel). For our scattering analysis, we focus on the energy-momentum region allowed for two chiral magnons localized at the lower edge, as highlighted by the blue shaded area in \figc{fig:3}{a}. The energy-momentum window $[E_{1}(K), E_{2}(K)]$ for such an asymptotic state of two edge magnons with momenta $k_1$ and $k_2$ is determined from the chiral magnon dispersion, i.e. $E_{1}(K)\!\leq\!\varepsilon(k_1) + \varepsilon(k_2)\!\leq\!E_{2}(K)$, with $k_1\!+\!k_2\!=\!K$, where we find scattering states labeled by the relative momenta $q=(k_1-k_2)/2$ and tabulate the scattering phase.
The scattering phase $\phi$ as a function of the relative momentum $q$ at different total momenta $K$ and anisotropies $\Delta$ is shown in \figc{fig:3}{b} for interchain hopping $J_{y}=0.5$.
As chiral magnons are strongly localized and thus predominately live on the outermost chain, it is pertinent to compare their scattering properties with those of magnons in a single one-dimensional XXZ chain (gray lines) \cite{takahashi2005thermodynamics}. 
Indeed, we find this to be a good approximation for most parameters, i.e. for small momenta $K \lesssim 0$ or large anisotropy $\Delta$. Interestingly, however, we find significant deviations of our numerical results from single chain analytical predictions for large momenta $K > 0$ and small anisotropies $\Delta\in\{0.1, 0.3\}$, indicating a strong renormalization of the interacting theory.
It is worth noting that we find real-valued $\phi$ only, confirming that there is indeed no weight of the wavefunction transferred to the bulk during scattering.\\
In \figc{fig:3}{c} we show the equal-momenta effective interaction $U_{\text{2-eff}}(K/2, K/2)$ as a function of the COM momentum $K$ for different anisotropies, observing strong deviations from the bare XXZ prediction for large and positive momenta. The observed strong renormalization effects in the edge theory can be traced back to bound modes in the bulk, which can resonate with the chiral magnon pair.\\
\section{Bulk Resonances} The energy of scattering edge states is determined by the one-magnon chiral band $\varepsilon(k)$. 
Depending on $\Delta$, different bulk modes can resonate with the scattering states. Two scattering magnons can form a metastable bound state: the longer the two magnons stay together, the larger their scattering shift will be. In principle, these metastable compounds could manifest for arbitrary velocities of the magnons, but we observe a strong signal only whenever their velocity is small. This suggests a hybridization with bulk bound states, which are not dispersive. It should be stressed that any bulk excitation has zero velocity in the thermodynamic limit, as semiclassically they move around cyclotron closed orbits with zero average velocity~\cite{landau1930, Landau1991}.
\begin{figure}
    \centering
    \includegraphics[width=\columnwidth]{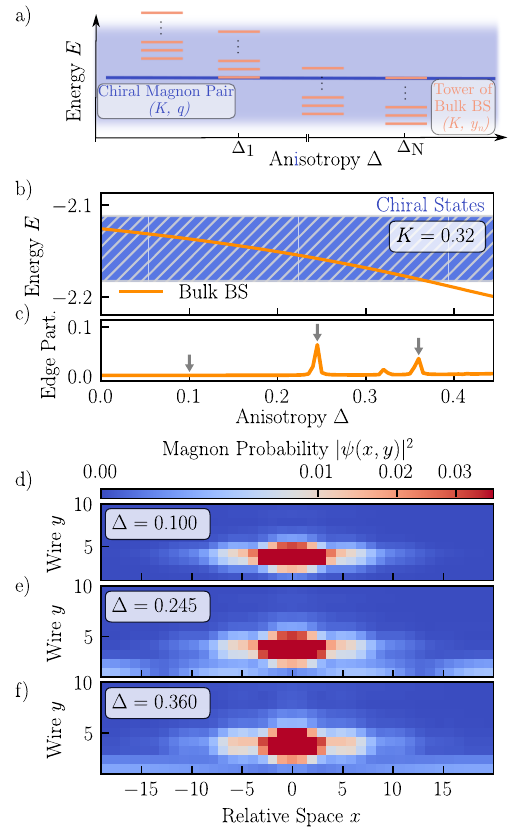}
    \caption{\textbf{Resonances.} a) Pictorial representation of scattering resonances. Upon changing $\Delta$, a set of bulk BS bands of total momentum $K$ gets tuned through a continuum of asymptotic states with variable relative momentum $q$, triggering resonances. b) Tracing one particular bound mode in the bulk of fixed total momentum $K=0.32$ through the continuum (blue region) of asymptotic states yields resonances with different chiral states. c) This is resolved in the participation of the wavefunction at the edge. For our finite size simulations we find three resonance peaks. d) - f) Resonances of the bulk BS with asymptotic states are reflected in the magnon distribution $\vert \psi(x,y) \vert^{2}$ in relative space $x$ and the bulk $y$. 
    In contrast to the unhybridized case d) the wavefunction contains strong contributions from the edge at the resonances (e,f). }
    \label{fig:4}
\end{figure}
Near the perturbative regime of weakly-coupled chains, each chain contributes one such strongly bound mode. Crucially, although the number of bound state bands scales with $L_{y}$, the number of resonances $N$ stays finite in the thermodynamic limit $L_{y}\to\infty$ since only bound modes located close enough to the boundary can overlap with edge modes.\\
The bound state energy is highly sensitive to changes in the anisotropy $\Delta$. As interactions increase, these modes are eventually pushed out of the two-particle continuum, forming stable bound states (see \App{app:ChiralBoundStates} for details)~\cite{Qin2017, Qin2018}.
For small $K\lesssim 0$, all relevant bulk modes are already situated lower in energy than the asymptotic energy window, even at $\Delta\!=\!0$, see \figc{fig:3}{a}, and so resonances are not possible. The situation is different though for large $K$.\\
The emergence of resonances between edge and bulk bound modes for varying $\Delta$ in the regime of large $K$ is entailed in hybridization of the associated wavefunctions shown in \figcs{fig:4}{b}{f}.
As $\Delta$ decreases from strong interactions, the energy of certain bulk-bound modes enters the two-chiral magnon continuum, allowing them to hybridize with edge states, see \figc{fig:4}{a}. This hybridization is detected through the edge participation factor $P\!=\!\sum_{x}\vert\psi(x,\!y\!=\!1)\vert^2$, which shows three distinct peaks at $\Delta\!\in\!\{0.245, 0.32, 0.36\}$, signaling strong coupling to different asymptotic scattering states, as shown in \figcc{fig:4}{b}{c}.\\ 
We then compare the unhybridized wave function, shown in \figc{fig:4}{d}, with the hybridized wave function at two of the most prominent resonances, as depicted in \figcc{fig:4}{e}{f}.  In all cases, the wave function remains sharply peaked at small relative distances $x$, but at resonance, a plane wave profile emerges in the magnon distribution near the edge, indicating hybridization with scattering states. The resonances change the effective magnon interactions, which confirms the behavior seen in \figc{fig:3}{c}. Notably, when comparing the hybridized cases in \figcc{fig:4}{e}{f}, additional nodes in the edge plane wave profile become apparent, which correspond to different relative momenta $q$ of the hybridizing asymptotic states.
As the system size increases along the $x$-direction, the spectrum of asymptotic edge states becomes denser, characterized by a continuous range of relative momenta $q$. Consequently, a discrete set of bulk modes is tuned through a continuum of edge modes. However, the extent of hybridization between bound bulk states and edge modes ultimately depends on the spatial overlap of their respective wave functions.\\
\section{Experimental Relevance} 
Signatures of this interplay between bulk and edge modes can also be examined in experiment. Inelastic STS allows systematic studies of scattering between chiral magnons via a real-time injection measurement protocol, see \figc{fig:1}{a}.
Our protocol consists of 3 steps: (i) Inject in the sample two magnon excitations of given energy $E(k_{1}), E(k_{2})$ within the bulk-gap via spin-exchange interactions between tip and material~\cite{ternes2015, Feldmeier2020}. Both magnons are injected with a time delay $t_{0}$ directly relating to the distance between both excitations as $s_{0}=v(k_{1}) t_0$, where $v(k_{1})$ denotes the velocity of the first excited slower magnon. (ii) After propagation both magnons will scatter after time $t^{*} = s_{0}/[v(k_{2}) - v(k_{1})]$ and experience scattering shifts $d_{1}(k_{1}, k_{2})$ and $d_{2}(k_{1}, k_{2})$, as illustrated in \figc{fig:1}{b}. (iii) After scattering both excitations are detected at time $t$ with distance $s(t) = [v(k_{2}) - v(k_{1})] (t - t^{*}) + [d_2(k_{1}, k_{2}) - d_1(k_{1}, k_{2})]$ using STM.
Hence monitoring the positions of both chiral magnons after scattering experimentally using local measurements and comparing to the free evolution of individual magnons allows us to detect $d_{1}, d_{2}$, see \figc{fig:1}{a}. 
Insights from this setup tunable in the energy and velocity of the injected magnons can later be applied to more complex transport experiments.
A more rigorous discussion of experimental caveats, including a comparison between the typically expected sizes of scattering shifts $d_{1}, d_{2}$ and experimental limitations due to measurement resolution of current STM setups, is given in \App{app:ExpDetails}. We emphasize that the most crucial requirement for the proposed experiment is the availability of a material realization of a topological magnon insulator with sufficiently long-lived edge states. While our construction scheme based on the AC effect enables a theoretical study of the relevant phenomena, experimental implementations are challenged by the weak coupling between spins and external electric fields. Consequently, strong electric field gradients are necessary, which may, in turn, induce additional intra-atomic effects.
However, a more promising route was recently proposed in \Reff{Nakata2017b}, where an array of STM tips is used to generate a skew-harmonic electromagnetic scalar potential, producing sizable electric-field gradients without the need for large overall electric-field amplitudes.\\

\section{Discussion} 
We have investigated the scattering of chiral edge magnons in a two-dimensional magnet subjected to an electric field gradient. By thoroughly analyzing the interactions among edge modes, we demonstrated their stability during scattering events and extracted their corresponding scattering phases, which fully captures the effective interactions.
It would be intriguing to use \eq{eq:eff_H_main} as a building block for a full-fledged many-body effective description, and take advantage of both powerful analytical methods
and advanced numerical techniques
available in one dimension.
For example, the long-wavelength dynamics of magnons with small relative momenta is that of one-dimensional bosons with contact interactions, i.e. the famous Lieb-Liniger integrable model \cite{Lieb1963, Lieb1963a}, whose thermodynamics \cite{takahashi2005thermodynamics} and hydrodynamics \cite{Bastianello2022} can be analytically studied.
Our study highlights the critical role of bulk spectra in shaping the many-body behavior of edge states in topological magnon systems. It would be exciting to see the phenomena we describe realized experimentally, for instance, through the proposed injection-measurement protocol. Promising platforms include topological magnon candidates like CrI$_3$ \cite{Chen2018CrI3}, where recent STM experiments indicate signatures of in-gap edge magnons \cite{zhang2024}, or artificial magnetic ad-atom arrays that mimic magnetic chains, where spin-orbit coupling can enhance the AC phase \cite{Katsura2005}.

\section*{Acknowledgments} 
We acknowledge support from the Deutsche Forschungsgemeinschaft (DFG, German Research Foundation) under Germany’s Excellence Strategy--EXC--2111--390814868, TRR 360 – 492547816, DFG grants No. KN1254/1-2, KN1254/2-1, and DFG Emmy Noether Programme---Project No. 504261060, the European Research Council (ERC) under the European Union’s Horizon 2020 research and innovation programme (grant agreement No. 851161), as well as the Munich Quantum Valley, which is supported by the Bavarian state government with funds from the Hightech Agenda Bayern Plus.\\
\section*{Data and Informations availability} 
Data, data analysis, and simulation codes are available upon reasonable request on Zenodo~\cite{zenodo}.

\begin{appendix}

\section{The Strong Anisotropy Limit -- Chiral Magnon Bound States}
\label{app:ChiralBoundStates}

In the main text we emphasized how strongly bound eigenstates can be tuned through resonances with individual asymptotic chiral states. Having increased the bare interaction strength $\Delta$ to sufficiently large values these strongly localized modes will transition to form stable bound modes, separated in energy from the two-particle continuum, as shown in \figc{fig:5}{a}. In total a number of $L_{y}$ bound state bands will experience this effect resembling the results for the single magnon spectrum. Each bound state band, however, is associated with twice the shift in momentum space. This is a result of the AC phase coupling to the physical spin charge of the excitation; the latter is double for the bound state compared to individual magnon excitations. In the spirit of the single magnon spectrum sufficient coupling between the wires, nonetheless, opens a bulk gap stabilizing two chiral bound state modes at the edges of the system, see \figcs{fig:5}{b}{d}. Same as the single magnon edge modes these edge bound states are stable throughout the tested parameter regime. This is consistent with findings from Ref.~\cite{Qin2017, Qin2018}.    

\section{Exact Diagonalization in Subspaces of Fixed Magnon Number}
\label{app:CharacterizationManyBodyEdge}

The conclusions drawn in our work are based on numerical results for single and two magnon spectra as well as characteristics of the corresponding wavefunctions. All presented results are obtained using exact diagonalization of the discussed Hamiltonian working either in a real space formulation or directly resolving the spectrum in momentum space. Our simulations thereby make use of magnon number conservation implied in Hamiltonian \eqref{eq:Hamiltonian} of the main text due to $U(1)$-symmetry around the axis of spin quantization $\hat{z}$. Exploiting this symmetry allows us to diagonalize sectors of a given magnon number individually. In contrast to an exponential scaling of the total Hilbert space dimension ($\propto2^{N}$) with the size of the system $N =L_{x}\times L_{y}$ these subsectors scale algebraically with the number of magnons, i.e. linear $\propto N$ for a single respectively quadratic $\propto N^{2}$ for two magnons. This beneficial scaling allows us to reach the system sizes required to study scattering between two magnons. For the extraction of scattering phases presented in \App{app:ExtractPhi} it is, moreover, required to obtain two-magnon eigenstates for a given total momentum $K$. For this, we assume periodic boundary conditions in the system along the wire direction $\hat{x}$ and derive the momentum space representation of $\mathcal{H}$. The latter is diagonal in the quantum number $K$ and can be diagonalized for each $K$ independently.

\section{Characterization of the Many-Body Theory of Edge Modes}
\label{sec:CharacterizationManyBodyEdge}
\subsection{Scattering Phases from Two Magnon Spectra}
\label{app:ExtractPhi}

\begin{figure*}
    \centering\includegraphics[width=\textwidth]{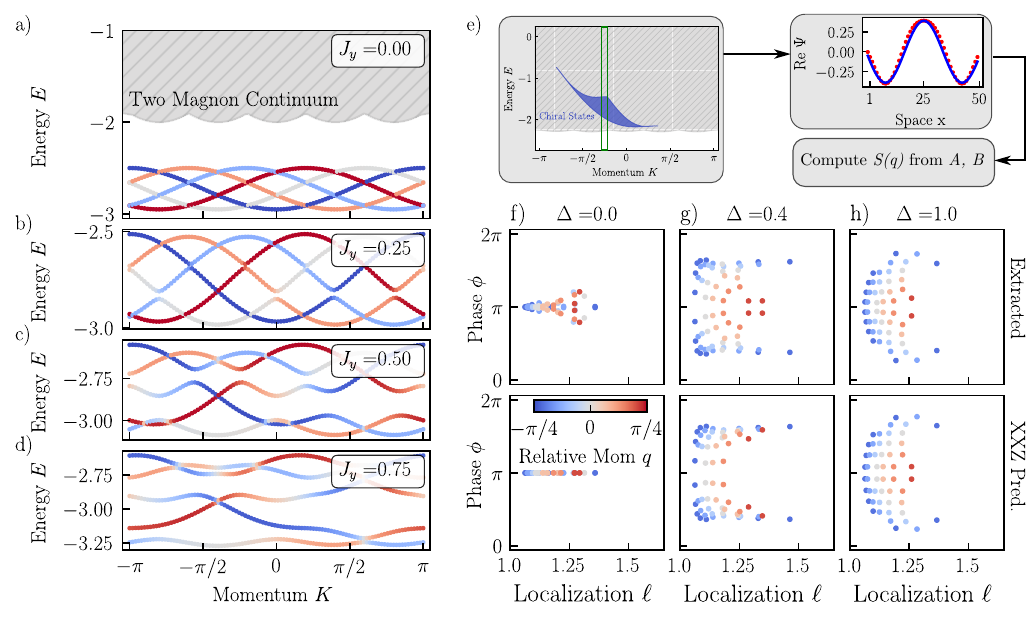}
    \caption{\textbf{Chiral Bound States and Scattering Phase Analysis.} a) Bound state bands separated in energy from the two magnon continuum for anisotropy $\Delta=1.25$ and uncoupled wires. b) - d) Increasing hopping strength between the wires $J_{y}\in \{0.25, 0.50, 0.75\}$ opens a bulk gap and stabilizes modes of chiral bound states. e) Analysis of the scattering phase consists of (1.) - (3.) selecting eigenstates within the chiral energy window for a given momentum sector $K$, (4.) fitting the states against a plane wave ansatz and (5.) compute the scattering matrix $S$. f) - h) Comparison of extracted (top row) and XXZ predicted (bottom row) scattering phases for different $K$ momenta and anisotropies $\Delta\in\{0.0, 0.4, 1.0\}$. The associated relative momenta are highlighted by color coding. Deviations are evident at low momenta ($\Delta=0.0$) and especially for less localized states.}
    \label{fig:5}
\end{figure*}

Our investigations on the edge theory in the main text are primarily based on the momentum dependent analysis of scattering phases. For this, we rely on a protocol consisting of the following points to extract scattering information from the numerically obtained spectra, also summarized in \figc{fig:5}{e}. 
\begin{enumerate}
    \item Determine the momentum-dependent energy window $[E_{1}(K), E_{2}(K)]$ possible for pairs of chiral magnons from the single magnon spectrum.
    \item Diagonalize the two magnon Hamiltonian in the corresponding total momentum sector $K$. 
    \item Filter the eigenspectrum for the eigenenergies within $[E_{1}(K), E_{2}(K)]$.
    \item Filter for asymptotic states described by a plane wave ansatz of the form $\psi(x, y_{1}, y_{2}) = A_{y_{1}, y_{2}}e^{iqx} +  B_{y_{1}, y_{2}} e^{-iqx}$ in the limit $x \gg 1$ for every pair of wire indices $(y_{1},y_{2})$, checking for agreement between various $y_{1}, y_{2}$. 
    \item From the fitted parameters $q, A, B$ compute the scattering matrix $S(q)=\frac{B}{A}$ as a function of relative momentum $q$.
\end{enumerate}
To increase the number of data points we additionally make use of properties under exchange of the scattering partners in the problem. Particle exchange results in a parity operation acting on the relative momentum $q \mapsto -q$, while leaving the center of mass momentum $K$ unaltered. The scattering element $S$ gets thereby mapped to its own inverse $S\mapsto S^{-1}$. We compare the extracted results against the analytical prediction for a XXZ spin chain taken from Ref.~\cite{takahashi2005thermodynamics},
\begin{gather}
\label{eq:XXZ}
    S_{\mathrm{XXZ}}(K, q) = e^{i\phi_{\mathrm{XXZ}}(K, q)} \nonumber \\  \phi_{\mathrm{XXZ}}(K, q) = \pi - i \log\Bigg[ \dfrac{1 - 4\Delta e^{i (\frac{K}{2} + q)} + e^{i K}}{1 - 4\Delta e^{i (\frac{K}{2} - q)} + e^{i K}}\Bigg].
\end{gather}

A comparison between the XXZ predicted (bottom row) and the numerically extracted (top row) scattering phases $\phi$ is shown in \figcs{fig:5}{f}{h} for anisotropies $\Delta\in[0.0, 0.4, 1.0]$. We collect scattering phases for values of $K\in[-\dfrac{\pi}{4}, \dfrac{\pi}{4}]$ and show the results as a function of the localization length $\ell$ of the eigenstates. To further discriminate the individual data points we highlight their relative momenta via color coding. The strongest deviations appear at small $\Delta$ for less localized states. This fits the semiclassical expectation of two more delocalized edge states having a reduced chance of scattering during their orbital motion at the boundary. In reality this enhanced leaking into the bulk, however, abets hybridization with bulk modes and the existence of resonances.\\
In the main text we also extract the effective interaction for the scattering phase data. For this, we determine the slope of $\phi(q)$ in the linear regime around relative momenta $q=0$. We correct the results by additionally weighting them with the effective magnon mass $m^{*}$ obtained from the curvature of the single magnon dispersion, i.e. $m^{*}(K) = \dfrac{1}{\partial_{K}^{2} \varepsilon(K)}$. The result quantifies the one-dimensional scattering problem yielding the inverse product of effective magnon mass and the scattering length $a$ for the edge theory, which is described in the next section. To quantify deviations from the bare XXZ theory we extract the same quantity from the analytical XXZ scattering prediction \eqref{eq:XXZ}
\begin{equation}
    c(K)  = \Bigg[-\frac{1}{m^{*}a}\Bigg](K) = \dfrac{4 \Delta \cos\big(K\big)}{2\Delta - \cos\bigg(\dfrac{K}{2}\bigg)}.
\end{equation}
The results summarize previous findings in a concise way; large (positive) momenta $K$ are associated with a strongly interacting edge theory while small (negative) momenta follow the bare XXZ prediction.  

\subsection{Comparison of Real and Fourier Space Results}

\begin{figure*}
    \centering
    \includegraphics[width=\linewidth]{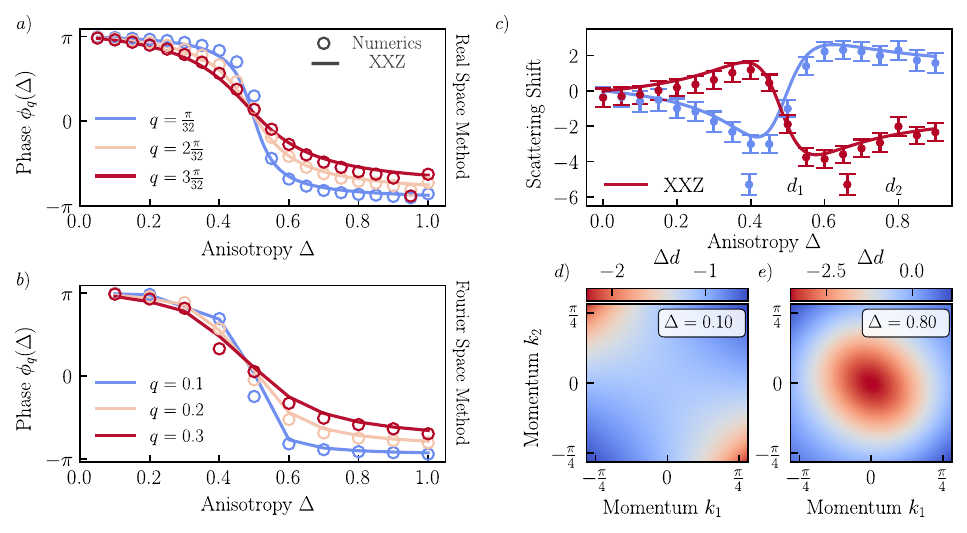}
    \caption{\textbf{Scattering Phases and Shifts from Real Space Simulations.} (a) We show scattering phases $\phi_{q}(\Delta)$ for two magnon scattering with total momentum $K=0$ and relative momenta $q\in\{\frac{\pi}{32}, \frac{2\pi}{32}, \frac{3\pi}{32}\}$ extracted from real space evolution. Results are set into comparison to XXZ predictions (solid lines). (b) Scattering phases $\phi_{q}(\Delta)$ for comparable parameters are extracted from the two-magnon spectra. The values of $q$ are thereby obtained from the fitting process and are approximately given by $q\in \{0.1, 0.2, 0.3\}$. (c) Scattering shifts $ d_{1},  d_{2}$ extracted from real space simulations of magnons pairs $(K, q) = (-\frac{\pi}{8}, \frac{\pi}{16})$ agree well with analytical predictions (solid lines) within the finite spatial resolution of the lattice. Error bars indicate the standard deviation of measurements of $d_{1},d_{2}$ at different times after the scattering event. (d)-(e) We show analytical predictions for the relative scattering shift $\Delta d=d_{1} -  d_{2}$ of two magnons following the XXZ description for anisotropies $\Delta=0.1$ (b) and $\Delta=0.8$ (c) in units of the lattice spacing $a$. We focus on the region of the Brillouin zone where chiral magnon bands are located, i.e. $k\in[-\frac{\pi}{4}, \frac{\pi}{4}]$.}
    \label{fig:6}
\end{figure*}

As emphasized before characterization of the effective theory of magnons at the edge is numerically most conveniently performed using the interacting eigenspectrum in momentum space. To relate to experiment, however, it is crucial to note that similar information can be obtained from time evolutions of real space scattering events. In experiment typically only the measured magnon position is available and the effective theory has to be deduced from analysis of the observed scattering shifts $\Delta d=d_{2}-d_{1}$. Numerical simulations of the real-time dynamics, however, allow us to access the complete two-magnon wavefunction, from which in particular the scattering phase $\phi$ can be determined. For our analysis this provides a natural way to compare the results obtained from Fourier and real space methods. To determine the scattering phase $\phi$ from real space evolutions we consider two magnon wavefunctions $\ket{\psi_{\Delta}(T)}$ at times $T$ after the scattering event has taken place. The evolution without explicit magnon interaction ($\Delta=0$) thereby serves as a reference state for evolutions with finite $\Delta$. More specifically the scattering phase $\phi(\Delta)$ is encoded in the overlap $\braket{\psi_{0}(T)}{\psi_{\Delta}(T)} \sim \exp[i \phi(\Delta) - i \varphi_{\text{dyn}}(T)]$. Besides the scattering phase $\phi(\Delta)$ generically also a dynamical phase contribution $\varphi_{\text{dyn}}(T)$ appears. The latter accounts for the phase two distant magnons accumulate under time evolution. For the formulation of the real space Hamiltonian of \eq{eq:Hamiltonian} in the main text it is given by $\varphi_{\text{dyn}}(T) = -\frac{J_{x}\Delta}{2} (L_{x}L_{y} - 8) T$. Results for the scattering phase $\phi_{q}(\Delta)$ from real space evolution of two magnons with vanishing total momentum ($K=0$) and relative momentum $q$ are shown in \figc{fig:6}{a}. We provide a comparison to analytical XXZ predictions (solid lines) and find good agreement for all $\Delta$ values. Applying the Fourier space method of \App{app:ExtractPhi} yields similar results; see \figc{fig:6}{b}. It is worth noting that while the relative momentum $q$ is fixed in real-space simulations of \figc{fig:6}{a} exactly by the definition of the initial wavefunction, see Eq.~(4) of the main text, the relative momenta $q$ contained in \figc{fig:6}{b} result from the fitting process of eigenstates explained \App{app:ExtractPhi} and hence are only approximately given by values of $q\in\{0.1, 0.2, 0.3\}$.\\
To furthermore connect to experiment, we can use the magnon profile of real space evolutions to directly determine the measured scattering shifts $ d_{1}, d_{2}$ of the first and second magnon, respectively. The extracted scattering shifts for a magnon pair of total momentum $K=0$ and relative momentum $q=\frac{3\pi}{32}$ are shown in \figc{fig:6}{c} and set into comparison to $d_{1} = \partial_{k_{1}}\phi_{\text{XXZ}}(k_{1}, k_{2})$  and $ d_{2} = \partial_{k_{2}}\phi_{\text{XXZ}}(k_{1}, k_{2})$ containing the XXZ scattering phase of \eq{eq:XXZ}. Indeed, we find characteristics of the analytical predictions clearly resolved in the experimentally accessible $d_{1}, d_{2}$. Interestingly, the scattering shifts $d_{1},d_{2}$ deviate more strongly from XXZ predictions than the phases $\phi_{q}(\Delta)$ of \figc{fig:6}{a}. This fact that can be attributed to the fitting process used to determine $ d_{1}, d_{2}$. For this, local peaks of the real space magnon distribution, indicating the magnon position, are extracted from simulations at multiple times after the scattering event and set into comparison with the free evolution of given magnon velocity. The standard deviation for measurements at different times is indicated by the error bars shown in \figc{fig:6}{c}. We find a finite spatial resolution of the real-space lattice, i.e., given by the lattice spacing, to be one crucial aspect limiting the accuracy of the extracted scattering.

\begin{figure*}
    \centering
    \includegraphics[width=\linewidth]{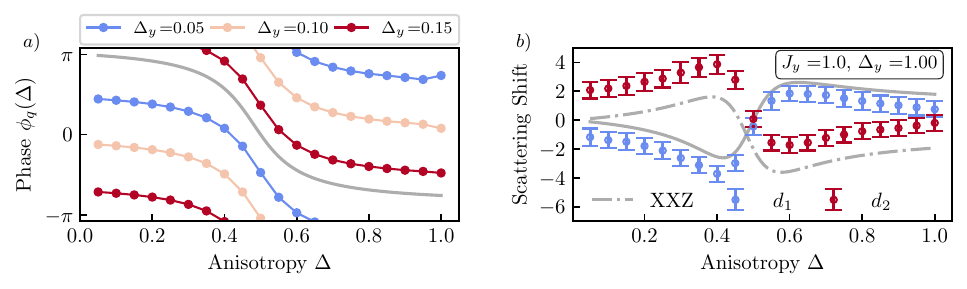}
    \caption{\textbf{Isotropic Magnon Interactions.} (a) Scattering phases $\phi$ for magnon scattering with momenta $(K, q) = (-\frac{\pi}{8}, \frac{\pi}{16})$ and interwire magnon interaction $\Delta_{y}\in \{0.05, 0.1, 0.15\}, J_{y}=0.5$ as a function of intrawire interaction $\Delta$. For comparison we show predictions for the XXZ scattering phase $\phi_{\mathrm{XXZ}}$ (gray line). (b) Scattering shifts extracted from real-time simulations for the isotropic system $\Delta_{y}=1.0, J_{y}=1.0$. Predictions for XXZ are plotted for comparison (gray lines).}
    \label{fig:7}
\end{figure*}

\section{Experimental Details}
\label{app:ExpDetails}

In our work, we propose an experimental protocol aiming to probe the scattering shift $\Delta d$ from interactions between two edge magnons using inelastic scanning tunneling microscopy (STM).
In the following, we want to judge whether the current experimental setups allow for a sufficiently high resolution to measure the relative scattering shift $\Delta d = d_1 - d_2$. Important contributions to this experimental resolution are (i) the width and (ii) the momentum spread of the magnon wavepackets injected in the beginning, and last but not least (iii) the spatial resolution $\Delta x$ of the detection. To estimate the magnitude of theses effects we perform valuations with parameters taken from current experimental platforms. As the width of the initialized wavepacket is expected to directly correlate with the typical range of interactions between tip and sample, which also determines the spatial resolution, we expect both effects (i) and (iii) to be on the same order of magnitude. For small sample-tip distances of $h$ ($\sim 0.5 - 1.5 \text{nm}$) and sharp tips of radius $R$ ($\sim 0.8\text{nm}$) the resolution can be computed within the Tersoff-Hamann model as $\Delta x=[(0.2~\text{nm}) (R + h)]^{\frac{1}{2}} \sim 0.5 \text{nm}$~\cite{bode2003a}. Next we have to analyze the momentum window $\Delta k$ excited due to finite energy resolution $\Delta E$ in STM. For the latter experiments working at high energy resolution demonstrated an accuracy of $\Delta E=0.1 \text{meV}$~\cite{wiebe2004, zhang2011, oka2014}. Assuming a more generous estimate of $\Delta E=0.25 \text{meV}$ and typical velocities of chiral magnons of $v\sim1000\frac{m}{s}$, e.g. for van-der-Waals magnets of lattice constant $a \sim 1\text{nm}$~\cite{neumann2024, zhu2021topological, Chen2021}, this relates to a momentum uncertainty of $\Delta k = \frac{\Delta E}{\hbar v} \sim 0.38 \frac{1}{\text{nm}}$. In conclusion the initialized wavepacket will spread over $6 \%$ of the Brillouin zone and potentially important broadening of the local wavepackets only appears if the dispersion shows non-negligible curvature over this range in momentum space. All effects combined hence cause an effect on the order of the lattice spacing $a$. What remains to be checked is if the scattering shift is expected to be resolvable taking into account the finite width of wavepackets and limited spatial resolution. For this we show analytical results for the XXZ scattering phase of \eq{eq:XXZ} for different interaction parameters $\Delta\in\{0.1, 0.8\}$ in \figcc{fig:6}{d}{e}. We find that the observed scattering shift $\Delta d$ typically takes values of a few lattice spacings $a$ and hence is within the resolution of current platforms.\\
Another aspect certainly relevant for experiment is the tunability of the bare interaction parameter $\Delta$. While in principle the latter can be tuned by applying external strain to the sample, there is an alternative way to cause resonances between edge modes and bulk bound modes in the system. Instead of tuning $\Delta$ variation of the Aharonov-Casher phase $\delta\varphi$ can lead to similar phenomena as demonstrated in the main text. Changes of $\delta\varphi$ do not shift the bands of bulk bound magnons in energy, but instead translate the momentum window of the chiral magnon band $\varepsilon(k)$ and with that move the energy-momentum window for two chiral edge magnons, see marked in blue in Fig.~3 a), within the Brillouin zone. 

\section{Isotropic Interactions}
\label{app:IsoInteractions}

While in the main text we discuss the limiting case of magnon interactions taking place only within the individual wires a more realistic scenario might include interactions also across different wires. To demonstrate that our analysis also applies in similar fashion to this scenario we generalize the coupling between chains and consider the Hamiltonian
\begin{widetext}
\vspace{-\baselineskip}
\begin{equation}
    \Ham = - \frac{J_{x}}{2}\sum_{x,y}\Bigg[\Big( e^{-i \varphi_{y}}\sigma_{x,y}^{+} \sigma_{x+1,y}^{-}  + \mathrm{h.c.} \Big) + \Delta \sigma_{x,y}^{z} \sigma_{x+1, y}^{z} \Bigg]  - \frac{J_{y}}{2}\sum_{x,y}\Bigg[\Big(\sigma_{x,y}^{+} \sigma_{x,y+1}^{-}  + \mathrm{h.c.} \Big) + \Delta_{y} \sigma_{x,y}^{z} \sigma_{x, y+1}^{z} \Bigg].
\end{equation}
\end{widetext}
In contrast to the previously discussed case, the presence of finite interactions $\Delta_{y}$ is expected to change the interacting theory on the edge already in the perturbative limit of weakly coupled chains ($J_{y}\ll 1$). This, in particular makes a comparison to analytical $XXZ$ results inapplicable. Scattering properties can, however, be analyzed following the same strategy as before. We show results for scattering phases $\phi$ for weak interwire interactions $\Delta_{y}\in\{0.05, 0.1, 0.15\}, J_{y}=0.5$ in \figc{fig:7}{a} and scattering shifts $d_{1}, d_{2}$ for isotropic coupling $J_{y}=1.0, \Delta_{y}=1.0$ in \figc{fig:7}{b}, respectively. Following the remarks of \Secc{ssec:EffectiveTheory} one can determine the effective interation strength at the edge. Relating the observed scattering properties to resonances with bulk modes would, however, require a separate analysis due to changes in the nature of bulk bound states due to finite interactions $\Delta_{y}$.

\end{appendix}
\let\oldaddcontentsline\addcontentsline
\renewcommand{\addcontentsline}[3]{}
\bibliography{TopologicalMagnons.bib}
\let\addcontentsline\oldaddcontentsline
\end{document}